\RequirePackage{fix-cm}
\documentclass[smallextended]{svjour3}       
\smartqed  
\usepackage{graphicx}
 \usepackage{natbib}

\RequirePackage{amsmath}
\usepackage{color}
\usepackage{graphicx}
\usepackage{booktabs, subfigure}
\usepackage{amssymb}
\usepackage{algorithm}
\usepackage{algpseudocode}

\newcommand{\vxi}{\mbox{\boldmath $\xi$}}

\newcommand{\vc}{\mathbf{c}}
\newcommand{\vY}{\mathbf{Y}}

\newcommand{\distras}

\newcommand{\bqa}{\begin{eqnarray*}}
\newcommand{\eqa}{\end{eqnarray*}}
\newcommand{\bqan}{\begin{eqnarray}}
\newcommand{\eqan}{\end{eqnarray}}
\newcommand{\bit}{\begin{itemize}}
\newcommand{\eit}{\end{itemize}}
\newcommand{\ben}{\begin{enumerate}}
\newcommand{\een}{\end{enumerate}}
\newcommand{\beq}{\begin{equation}}
\newcommand{\eeq}{\end{equation}}
\newcommand{\bdes}{\begin{description}}
\newcommand{\edes}{\end{description}}

\begin{document}

\title{Selection of the Number of Clusters in Functional Data Analysis\\
Adriano Zanin Zambom - Julian A. Collazos - Ronaldo Dias
}


\author{Adriano Zanin Zambom         \and
        Julian A. A. Collazos \and Ronaldo Dias 
}
\author{}


\institute{A. Z. Zambom \at
              Department of Mathematics, California State University Northridge, USA \\
              Tel.: +8186772701              \\
              \email{adriano.zambom@csun.edu}           
           \and
           J. A. A. Collazos \at New Granada Military University\\
              \email{jualacco@gmail.com}           
           \and
           R. Dias \at State University of Campinas\\
              \email{dias@ime.unicamp.br}           
           }

\date{Received: date / Accepted: date}

\maketitle

\begin{abstract}
Identifying the number $K$ of clusters in a dataset is one of the most difficult problems in clustering analysis. A choice of $K$ that correctly characterizes the features of the data is essential for building meaningful clusters. In this paper we tackle the problem of estimating the number of clusters in functional data analysis by introducing a new measure that can be used with different procedures in selecting the optimal $K$. The main idea is to use a combination of two test statistics, which measure the lack of parallelism and the mean distance between curves, to compute criteria such as the within and between cluster sum of squares. 
Simulations in challenging scenarios suggest that procedures using this measure can detect the correct number of clusters more frequently than existing methods in the literature. The application of the proposed method is illustrated on several real datasets.
\end{abstract}

\section{Introduction}
\label{intro}

Various methods have been proposed for clustering items in multidimensional spaces. The identification of homogeneous subsets of the data is a fundamental task in diverse statistical applications such as gene expression data (\cite{SorlieEtAl2003}, \cite{ma2006data}, \cite{NairEtAl2019}), finance (\cite{WangEtAl2006}, \cite{LascioEtAl2018}), medicine (\cite{McClellandKronmal2002}, \cite{KimEtAl2014}, \cite{ZhangEtAl2017}), just to cite a few. Selecting the number $K$ of clusters is an essential problem, so that objects in the same group have  features as similar as possible, and objects in different groups have distinct features. In general, the number of clusters in a dataset is unknown, however in some cases, the choice can be motivated by the dataset itself. An area where finding the number of clusters and the clusters themselves can be especially challenging is functional data analysis, which requires working with spaces of infinite dimension. 
In this paper we introduce a new measure that can be used with several criteria to select the number of clusters in functional data analysis.

 Several authors in the literature of statistics have considered the problem of selecting the number of clusters in a dataset. The many criteria for finding the best (in some sense) number of clusters are a general guideline, however the researcher has to decide the appropriate definition of ``clusters", or the intended definition of separation between clusters, for each specific application (\cite{Hennig2015}).
A comparison of early methods in the general multidimensional setting can be found in \cite{milligan1985examination} and \cite{Hardy1996}. The well-known distance-based clustering algorithms, such as the $K$-means, quantify the closeness between observations by some distance measure.
One of the classical approaches is the \cite{calinski1974dendrite} dendrite method, which selects the number of clusters that maximizes the ratio of the between to the within cluster sums of squares. \cite{krzanowski1985criterion} suggest choosing $K$ to maximize $DIFF(k)/DIFF(k+1)$, where $DIFF(k)$ is a scaled difference between the within sum of squares when using $k-1$ and that when using $k$ clusters. Similarly, \cite{hartigan1975clust} proposes minimizing a sequential measure of ratios between the within sums of squares.
\cite{tibshirani2001estimating} introduce the interesting Gap statistic, which uses the output of any clustering algorithm and compares the expected within-cluster dispersion of the null distribution and the one computed from the data.
 Other methods include  \cite{kaufman2009finding}, \cite{fischer2011number}, \cite{Steinwart2015}, \cite{HyrienBaran2016}, \cite{ChakrabortyDas2018}, and references therein.
 In general, methods either perform clustering for different values of $k$ and select the one that minimizes/maximizes some criterium, or decide at each step whether a cluster should be split into two clusters.



Clustering functional data is not new in the statistics literature, see for instance \cite{AbrahamEtAl2003}, \cite{TokushigeEtAl2007}, \cite{BouveyronJacques2011},   \cite{Febrero-BandeFuente2012}, \cite{IevaEtAl2013}, \cite{jacques2013funclust} \cite{YamamotoTerada2014}, \cite{WangEtAl2014}, \cite{JacquesPreda2014}, \cite{GarciaEtAl2015}, \cite{bouveyron2015discriminative} and references therein. Model-based clustering techniques, which assume that the data comes from a mixture of probability distributions, usually select the  number of clusters using the Bayesian Information Criteria (BIC). For instance, \cite{ma2006data} and \cite{ma2008penalized} use the BIC while performing clustering via the rejection-controlled EM-algorithm initialized by $K$-means clustering for functional Gaussian mixture models.  \cite{bouveyron2011model} study clustering of time series in group-specific functional subspaces, while \cite{giacofci2013wavelet} develop a clustering method for  mixed-effects functional models in high dimensions, both using the BIC to select the number of clusters. Other authors who perform clustering in functional data analysis and use specially designed methods to select the number $K$ of clusters
 include \cite{serban2012multilevel} in a multilevel functional  clustering; 
\cite{james2003clustering}, who use the largest jump in distortion function, which is defined as the average Mahalanobis distance between each spline coefficient in the functional clustering model and its closest cluster center; and
 \cite{ieva2013multivariate}, who choose the optimal number of clusters using the so-called silhouette (\cite{kaufman2009finding}) values.


A basic framework to select the number of clusters in functional datasets is to use well-established procedures with some distance measure between curves (and distance between curves and cluster centers). These distances can be used, for example, to compute the within and between cluster sum of squares (\cite{calinski1974dendrite}, \cite{krzanowski1985criterion}, \cite{hartigan1975clust}).
In functional data analysis, the distance between curves can be computed as the square root of the sum of the $L_2$ distances at each observed time point. However, the $L_2$ distance may not always be the best measure when clustering functional data. In this paper a new measure of similarity of curves is introduced, which combines two test statistics that assess the parallelism and the mean height of each curve. The test of parallelism is a novel concept adapted from \cite{ZambomAkritas2014} and \cite{zambom2018functional}. 
The proposed measure is then used to alter some existing approaches that attempt to minimize a distance-based dissimilarity measure of within and between clusters as well as other criteria. The choice of the number of clusters using the proposed measure seems to enhance the stability and agreement of the modified procedures when compared to the same procedures using the $L_2$ version and those methods specially designed for functional data.

The remainder of the paper is organized as follows. Section \ref{methodnumclust} presents the proposed measure and how it can be used to adapt existing methods to select the number of clusters. Section \ref{simnumclust5} shows a comparative numerical simulation of the performance of the proposed method and other existing approaches in the literature, including some that use BIC or AIC. A data analysis of real datasets is presented in Section \ref{realdatanumclust5}, and Section \ref{conclus5} provides some concluding remarks.

\section{Methodology}\label{methodnumclust}

\indent \indent Let $X_\ell^i(t), t \in \mathcal{T}$ denote the true functional data of the $i$-th observational unit, $i = 1, \ldots, n$, in cluster $\ell$, $\ell = 1, \ldots, K$, where $\mathcal{T}$ is a compact set in $\mathbb{R}$, usually representing time. Since functional data are in general discretely recorded and contaminated with measurement errors, let 
\bqan\label{model}
Y^{i}(t_{j}^i) = X_\ell^{i}(t_{j}) + \varepsilon_{j}^{i}, \;\;\; j = 1, \ldots, r,
\eqan
be the observed values of the $i$-th function $X_\ell^i()$ at  $t_1, \ldots, t_{r}$, where $\varepsilon_{j}^{i}$ is the independent and identically distributed measurement error.  
This is a general formulation of functional data arising from clusters, which includes for example the model in \cite{ma2006data} who uses $X_\ell^{i}(t_{j}) = \mu_\ell(t_{ij}) + b_i$, where $\mu_\ell()$ is the mean curve of cluster $k$ and $b_i$ is a Gaussian parallel deviation. The cluster centers in this case are the mean functions $\mu_\ell()$. For a more general notation we will denote the center of the $\ell$-th cluster by $\vc^{\ell} = (c^\ell(t_1), \ldots,c^\ell(t_r)) , \ell = 1, \ldots, K$. Assuming that the true data generating mechanism is the one in model (\ref{model}) but the curves are observed without the cluster labels, the objective is to correctly identify $K$, the true number of clusters.

 The proposed measure to be used in selecting the number of clusters is based on a combination of two test statistics, which we derive using ideas similar to those in \cite{zambom2018functional}. These test statistics evaluate the parallelism and average height between curves, so that our method aims at identifying clusters that are determined by the curve shape and height.
The first statistic checks whether the curve of an individual is parallel to a cluster center in the following way. Compute the residuals
\bqa
\xi^{i\ell}_j = Y^i(t_j) - c^\ell(t_j), \hspace{.5cm} j = 1, \ldots, r, \;\; i = 1, \ldots, n, \;\; \ell = 1, \ldots, K.
\eqa
Consider $(\xi^{i\ell}_j, t_j)$ as data from a one-way ANOVA design with $\xi^{i\ell}_j$ being the observation at ``level" $t_j$. Augment each ANOVA factor level $t_j$ by including the $\xi^{i\ell}_j$ corresponding to the $(m-1)$, for $m$ odd, nearest grid points on either side of $t_j$, that is
\bqa
W_{j} = \left\lbrace s: |t_j - t_s|\leq \frac{m-1}{2r}\right\rbrace.
\eqa
The choice of $m$ determines the size of each factor level in the augmented ANOVA, and can be chosen by running the test after randomly
permuting the observed variables among the times $t_j$, in order
to induce the validity of the null hypothesis, and selecting $m$ to approximate the desired level of the test (\cite{ZambomAkritas2014}, \cite{ZambomAkritas2015}, \cite{ZambomKim2017}).

  If the curve is parallel to the cluster center, the residuals $\xi^{i\ell}_j$ should have constant mean and thus a test for parallelism is equivalent to $H_0: E(\xi^{i\ell}_j) = a$, for a constant $a$. This test can be accomplished with an ANOVA F test of constant mean across the factor levels defined by $W_{j}$. Hence we propose using the one-way ANOVA test statistic 
\bqa
MST(\vxi^{i\ell})-MSE(\vxi^{i\ell}) = \frac{m}{r-1}\displaystyle\sum_{j=1}^{r}(\xi^{i\ell}_{j\cdot} - \xi^{i\ell}_{\cdot\cdot})^{2} - \frac{1}{r(m-1)}\displaystyle\sum_{j=1}^{r}\displaystyle\sum_{s\in W_{j}}^{m}(\xi^{i\ell}_{s}-\xi^{i\ell}_{j\cdot})^2,
\eqa
where $\vxi^{i\ell} = (\xi^{i\ell}_1, \ldots, \xi^{i\ell}_r)$, $\xi^{i\ell}_{j\cdot}=(1/m)\sum_{s\in W_{j}}\xi^{i\ell}_s$, $\xi^{i\ell}_{\cdot\cdot} = (1/rm)\sum_{j=1}^{r}\sum_{s\in W_{j}}\xi^{i\ell}_{s}$, and $MST(\vxi^{i\ell})$ and $MSE(\vxi^{i\ell})$ are the treatment mean sum of squares and error mean sum of squares, respectively, from an ANOVA whose factor levels are defined  by the windows $W_j$. The form of this statistic is due to \cite{AkritasPapadatos2004}, who showed that its distribution is asymptotically equivalent to the distribution of the well-known $F =MST(\vxi^{i\ell})/MSE(\vxi^{i\ell})$ when the number of factor levels goes to infinity.
In order to obtain a non-negative measure, we compute the standardized version
\bqa
T(\vxi^{i\ell}) = \left| \frac{\sqrt{n}\left( MST(\vxi^{i\ell}) - MSE(\vxi^{i\ell})\right)}{\widehat{\tau}_{i\ell}\sqrt{2m(2m-1)/(3(m-1))}}\right|,
\eqa
where 
\bqa
\widehat{\tau}_{i\ell}^{2}=\frac{1}{4(n-3)}\displaystyle\sum_{s=2}^{r-2}(\xi^{i\ell}_s-\xi^{i\ell}_{s-1})^{2}(\xi^{i\ell}_{s+2}-\xi^{i\ell}_{s+1})^{2}
\eqa
is the estimated standard deviation of the test statistic. Large values of $T(\vxi^{i\ell})$ indicate low probability (small p-value) that the residuals have constant mean, which suggests that the curve is not parallel to that cluster center, and hence should not be assigned to that cluster.

The second statistic is based on the t-test for differences in averages, which is defined as
\bqa
W(\vY^{i}, \vc^{\ell}) = \left| \frac{\frac{1}{r}\sum_{j=1}^{r} Y^i(t_j) - \frac{1}{r}\sum_{j=1}^{r}c^\ell(t_j)}{\sqrt{\frac{\widehat{V}(\vY^i)+\widehat{V}(\vc^\ell)}{r}}}\right|,
\eqa
 where $\vY^{i} = (Y^{i}(t_1), \ldots, Y^{i}(t_{r}))$,  $\vc^{\ell} = (c^{\ell}(t_1), \ldots, c^{\ell}(t_{r})$, and $\widehat{V}$ is an unbiased estimator of their variance. Again, large values of $W(\vY^{i}, \vc^{\ell})$ suggest that the curve does not belong to that cluster.

Because in practice there is more than one possible definition of clusters, and in fact the true underlying number of clusters may not even exist, the choice of method to determine the best number of clusters $K$ should depend on the application at hand and the researcher's definition of clusters and expected separation.
The proposed method can be used to modify a variety of procedures in the literature to select the number of clusters, hence being adaptive to different requirements one may have when defining clusters and their meaningful structure in each application.
The adaptation of some procedures in the literature with the aforementioned test statistics can be performed in the following way.
Define the measure that combines the two test statistics by
\bqa
TW_{\alpha}(\vxi^{i\ell}, \vY^{i}, \vc^{\ell}) = \sqrt{\alpha T(\vxi^{i\ell}) + (1-\alpha)W(\vY^{i}, \vc^{\ell})}.
\eqa
The tuning parameter $\alpha$ determines how much weight should be given to the measure of parallelism and  how much weight should be given to the average height distance between curves and cluster centers (see Section \ref{select.alpha} for a discussion on how to select $\alpha$).
Define the within-cluster sums as
\begin{equation}\label{WSS}
WCS_\alpha(k)=\displaystyle\sum_{\ell=1}^{k}\displaystyle\sum_{i\in S^{\ell}}TW_{\alpha}\left(\vxi^{i\ell}, \vY^{i}, \vc^{\ell}\right),
\end{equation}
 and the between-cluster sums as
\begin{equation}\label{BSS}
BCS_\alpha(k)=\displaystyle\sum_{\ell=1}^{k}n_{\ell}TW_{\alpha}\left(\vc^\ell-\vc, \vc^{\ell}, \vc\right),
\end{equation}  
where $\vc$ is the cluster center for all curves and $n_{\ell}$ is the number of curves in the $\ell$-th cluster. 
Using the between and within sums  defined in Equations (\ref{WSS}) and (\ref{BSS}), we modify the following methods for choosing the optimal number of clusters.

\textbf{Calinski and Harabasz} (\cite{calinski1974dendrite}) Let
\begin{equation}\label{CH}
CH_\alpha(k)=\dfrac{BCS_\alpha(k)/(k-1)}{WCS_\alpha(k)/(n-k)}
\end{equation}
and choose
$$K_{opt} = \text{arg}\max_{k}\:CH_\alpha(k).$$

\textbf{Krzanowski and Lai} (\cite{krzanowski1985criterion}) Let
\begin{equation}\label{KL}
KL_\alpha(k)=\left|\dfrac{DIFF_\alpha(k)}{DIFF_\alpha(k+1)}\right|,
\end{equation}
where  
$$DIFF_\alpha(k)=(k-1)^{2/r}WCS_\alpha(k-1)-k^{2/r}WCS_\alpha(k),$$
and choose
$$K_{opt} = \text{arg}\max_{k}\:KL_\alpha(k).$$

\textbf{Hartigan} (\cite{hartigan1975clust}) Let
\begin{equation}\label{H}
H_\alpha(k)=(n-k-1)\left[\dfrac{WCS_\alpha(k)}{WCS_\alpha(k+1)}-1\right]
\end{equation}
and choose
$$K_{opt} = \text{arg}\min_{k}\:\lbrace H_\alpha(k)\leq 10\rbrace.$$

\textbf{Kaufman and Rousseeuw} (\cite{kaufman2009finding}) Define the Silhouette statistic as
\begin{equation}\label{S}
S_\alpha(k)=\frac{1}{n}\displaystyle\sum_{i=1}^{n}\dfrac{b_\alpha(i)-a_\alpha(i)}{max\lbrace a_\alpha(i),b_\alpha(i)\rbrace}
\end{equation}
and choose
$$K_{opt} = \text{arg}\:\max_{k}\:S_\alpha(k),$$
where $a_\alpha(i)$ is the average of $TW_{\alpha}(\vY^{i} - \vY^{j}, \vY^{i}, \vY^{j})$ for $j = 1, \ldots, n_\ell$ corresponding to the data in the same cluster as curve $i$; and  $b_\alpha(i)$ is the average of $TW_{\alpha}(\vY^{i} - \vY^{j}, \vY^{i}, \vY^{j})$ for curves in the nearest cluster, where nearest is defined as the cluster minimizing $b(i)$.\\

\textbf{Tibshirani, Walther and Hastie} (\cite{tibshirani2001estimating}) Define the Gap statistic as
\begin{equation}\label{gap} 
Gap_\alpha(k)=\dfrac{1}{B}\sum_{b=1}^{B}log(\mathcal{W}_\alpha^{\ast b}(k))-log(\mathcal{W}_\alpha(k)),
\end{equation}
where 
$$
\mathcal{W}_\alpha(k)=\displaystyle\sum_{\ell=1}^{k}(2n_{\ell})^{-1}\displaystyle\sum_{i\in c^{\ell}}TW_{\alpha}\left(\vxi^{i\ell}, \vY^{i}, \vc^{\ell}\right).
$$
Here, $B$ different uniform data sets, each with the same range as the original data, are generated and the within cluster sum is calculated for different numbers of clusters. $\mathcal{W}_\alpha^{\ast b}(k)$ is the within cluster sum for the $b$-th simulated uniform data set. One approach would be to maximize $Gap_\alpha(k)$. However, to avoid adding unnecessary clusters an estimate of the standard deviation of $log(\mathcal{W}_\alpha^{\ast b}(k))$, $s_{k}$, is computed and the optimal choice is
$$K_{opt} = \text{arg}\:\min_{k}\:\lbrace Gap_\alpha(k)\geq Gap_\alpha(k+1)-s_{k+1}\rbrace.$$

\textbf{James and Sugar} (\cite{james2003clustering}) Use the distortion function defined as
\begin{equation}\label{JUMP}
d_\alpha(k)=r^{-1}\min_{c^{1},..., c^{k}}E\left[WCS_\alpha(k) \right].
\end{equation}
 The distortion $d_\alpha(k)$ is the average measure between each curve $i$ and its closest cluster center $\ell$, for $\ell = 1, \ldots, k$. Choose
$$K_{opt} = \text{arg}\max_{k}\:\lbrace d_\alpha^{-t}(k)-d_\alpha^{-t}(k-1)\rbrace,$$

with $t=r/2$.

\textbf{Fischer} (\cite{fischer2011number}). Define 
\begin{equation}\label{DDSEJUMP}
\delta_\alpha(k, \lambda)= \text{arg}\:\min_{k}\left[\Lambda_\alpha(\vY, \widehat{\textbf{c}}^{k})+ \lambda\sqrt{k/n}\right],
\end{equation}\\
where $\widehat{\textbf{c}}^{k}$ is a minimizer of the criterium $\Lambda_\alpha(\vY,\textbf{c})$ over $\Gamma_{k}$, where $\Gamma_k$ is the set of $(c_1, \ldots, c_k)$, that is, $\widehat{\textbf{c}}^{k}\in \:\: \text{arg min}_{\textbf{c}\in\Gamma_{k}} \Lambda_\alpha(\vY, \textbf{c})$. The first term for every $k$, is defined as the empirical distortion
\bqa
\Lambda_\alpha(\vY,\textbf{c})=\dfrac{1}{n}\sum_{i=1}^n \min_{\ell = 1\ldots k} TW_{\alpha}(\vxi^{i\ell}, \vY^{i}, \vc^{\ell}).
\eqa
The second term in Equation (\ref{DDSEJUMP}) is a first approximation for the penalty shape which depends on $k$ and tends to 0 at the rate $1/\sqrt{n}$.
To determine the constant $\lambda$, the slope heuristics (\cite{birge2007minimal}) is used. 
The penalty can be calibrated by two methods: the Jump method (\textbf{Djump}), which consists in identifying an abrupt jump in the model complexity, and the DDSE (\textbf{DDSE}) method, which computes the slope of the line of the empirical contrast and the penalty shape for complex models. Both methods have been implemented in the R (www.r-project.org) package \textit{capushe} (CAlibrating Penalty Using Slope HEuristics) by \cite{baudry2012slope}.

\subsection{Choice of $\alpha$} \label{select.alpha}

The parameter $\alpha$ determines the weight given to the parallelism of the curves or the difference in mean values of the curves. It is convenient to design an automatic procedure for choosing the weight $\alpha$, so that one does not have to rely on ad hoc choices that can differ from dataset to dataset and from one procedure to another.  Note that by modifying the methods as described in Section \ref{methodnumclust} with the dependence in $TW_\alpha$, each method depends on $\alpha$ and so does each selected $K_{opt}$, however, this was omitted from the description for ease of notation.

Because each method of selecting $K$ is computed using different combinations of $WCS$, $BCS$, and other measures, the choice of $\alpha$ may depend on the characteristics of each criteria and how they are used. As such, we propose selecting $\alpha$ and $K$ for each method by optimizing the criteria concomitantly on $k$ and $\alpha$. For example, when using Calinski and Harabasz's procedure, select the optimal $K$ by computing
$$\begin{pmatrix} K_{opt}  \\ \alpha \end{pmatrix}  = \text{arg}\max_{k,\alpha}\:CH_\alpha(k),$$
and selecting $K_{opt} = e^T\begin{pmatrix} K_{opt}  \\ \alpha \end{pmatrix}$
where $e^T = (1,0)$. Similar arguments can be used to select the optimal $K$ with optimal $\alpha$ for all the methods considered in this paper.

\section{Numerical simulations}\label{simnumclust5}

In this section we investigate the finite sample performance of the proposed method in selecting the number of clusters through several simulation scenarios. For comparison purposes, we evaluate the performance of the selection procedures using the proposed measure $TW_\alpha(\cdot)$ as well as the $L_2$ distance. We also include the results of some competing methods in the literature that are designed specifically for functional data, namely funclust (\cite{jacques2013funclust}), funHDDC (\cite{bouveyron2011model}), curvclust (\cite{giacofci2013wavelet}), funFEM (\cite{bouveyron2015discriminative}), and MFclust (\cite{ma2006data}). These methods are readily available in the statistical software R.

We consider 4 scenarios with varying degrees of difficulty. For each cluster $\ell = 1, \ldots, K$ in each scenario, $n_\ell$ curves are generated as follows
\bqa
&&\textit{Scenario 1}: \text{K = 2 clusters}\\   
&&\hspace{1cm} \text{Cluster 1}: a_i + 1.8 - \cos(1.8\pi x ) - \sin(1.8x) + \epsilon_i, i = 1, \ldots,  43 = n_1,\\
&&\hspace{1cm} \text{Cluster 2}: a_i +  2.4 - \cos(1.8\pi x ) - \sin(2.2x) + \epsilon_i, i = 1, \ldots,  57 = n_2,\\
&&\hspace{1cm}\mbox{where } a_i \sim U(-1/3, 1/3), \epsilon_{i} \sim N(2, 0.4^2).\\
&&\textit{Scenario 2}: \text{K = 3 clusters}\\   
&&\hspace{1cm} \text{Cluster 1}: a_i + 2.0 - \cos(1.5\pi x^3) + \epsilon_i, i = 1, \ldots, 35 = n_1,\\
&&\hspace{1cm} \text{Cluster 2}: a_i +  2.3 - \cos(1.7\pi x^3 ) + \epsilon_i, i = 1, \ldots,  43 = n_2,\\
&&\hspace{1cm} \text{Cluster 3}: a_i +  2.7 - \cos(1.9\pi x^3 ) + \epsilon_i, i = 1, \ldots,  72 = n_3,\\
&&\hspace{1cm}\mbox{where } a_i \sim U(-1/2, 1/2), \epsilon_{i} \sim N(2, 0.4^2).\\
&&\textit{Scenario 3}: \text{K = 4 clusters}\\   
&&\hspace{1cm} \text{Cluster 1}: a_i + 0.9 + \sin(1.5\pi x) + \sin(\pi x^2) + \epsilon_i, i = 1, \ldots, 34 = n_1,\\
&&\hspace{1cm} \text{Cluster 2}: a_i +  1.5 + \sin(1.7\pi x) + \sin(\pi x^2) + \epsilon_i, i = 1, \ldots,  67 = n_2,\\
&&\hspace{1cm} \text{Cluster 3}: a_i +   2.2 + \sin(1.9\pi x) + \sin(\pi x^2) + \epsilon_i, i = 1, \ldots,  71 = n_3,\\
&&\hspace{1cm} \text{Cluster 3}: a_i +   2.4 + \sin(1.6\pi x) + \sin(\pi x^2) + \epsilon_i, i = 1, \ldots,  28 = n_4,\\
&&\hspace{1cm}\mbox{where } a_i \sim U(-1/3, 1/3), \epsilon_{i} \sim N(2, 0.4^2).\\
&&\textit{Scenario 4}: \text{K = 5 clusters}\\   
&&\hspace{1cm} \text{Cluster 1}: a_i +\cos(\pi x) - x^2 + \epsilon_i, i = 1, \ldots, 50 = n_1,\\
&&\hspace{1cm} \text{Cluster 2}: a_i +\cos(1.2\pi x) - x^2 + \epsilon_i, i = 1, \ldots, 62 = n_2,\\
&&\hspace{1cm} \text{Cluster 3}: a_i +\cos(1.4\pi x) - x^2 + \epsilon_i, i = 1, \ldots, 36 = n_3,\\
&&\hspace{1cm} \text{Cluster 4}: a_i +\cos(1.6\pi x) - x^2 + \epsilon_i, i = 1, \ldots, 43 = n_4,\\
&&\hspace{1cm} \text{Cluster 5}: a_i +\cos(1.8\pi x) - x^2 + \epsilon_i, i = 1, \ldots, 59 = n_5,\\
&&\hspace{1cm}\mbox{where } a_i \sim U(-1/4, 1/4), \epsilon_{i} \sim N(2, 0.3^2).\\
\eqa

\begin{figure}[htbp!]
     \begin{center}
        \subfigure[]{%
            \label{fig:first}
            \includegraphics[width=5cm,height=3cm]{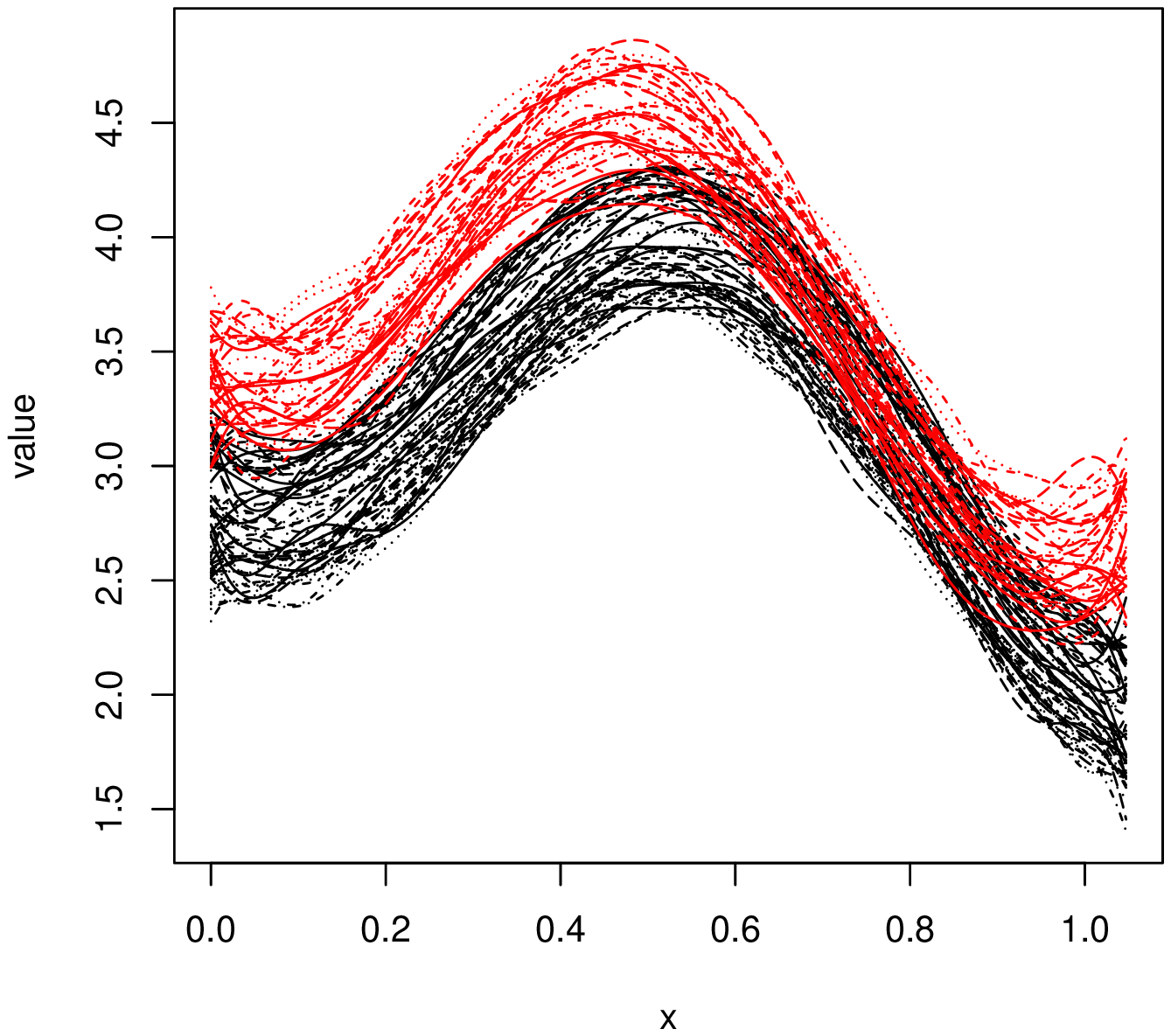}
        }
        \subfigure[]{%
           \label{fig:second}
           \includegraphics[width=5cm,height=3cm]{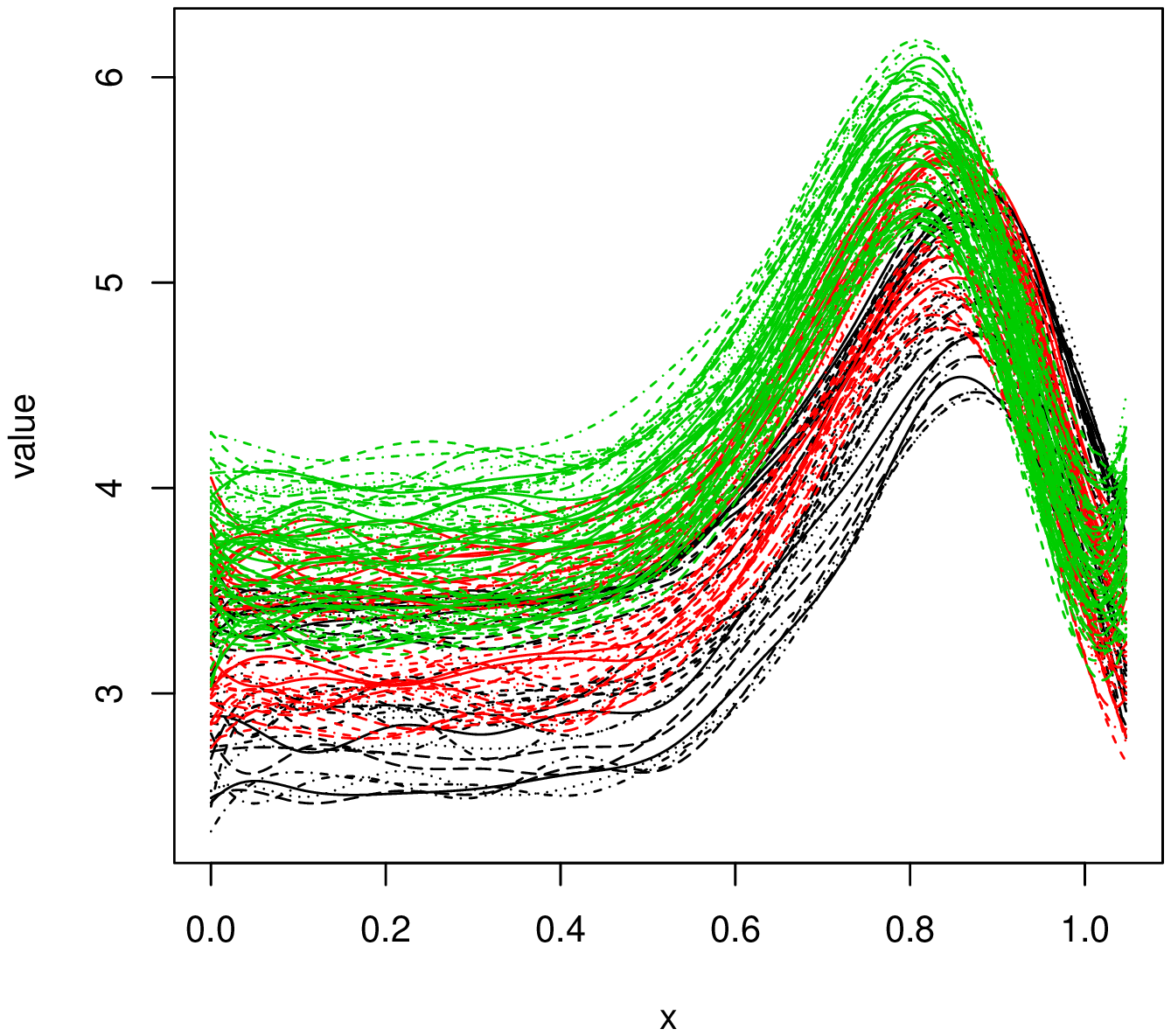}
        }\\ 
        \subfigure[]{%
            \label{fig:third}
            \includegraphics[width=5cm,height=3cm]{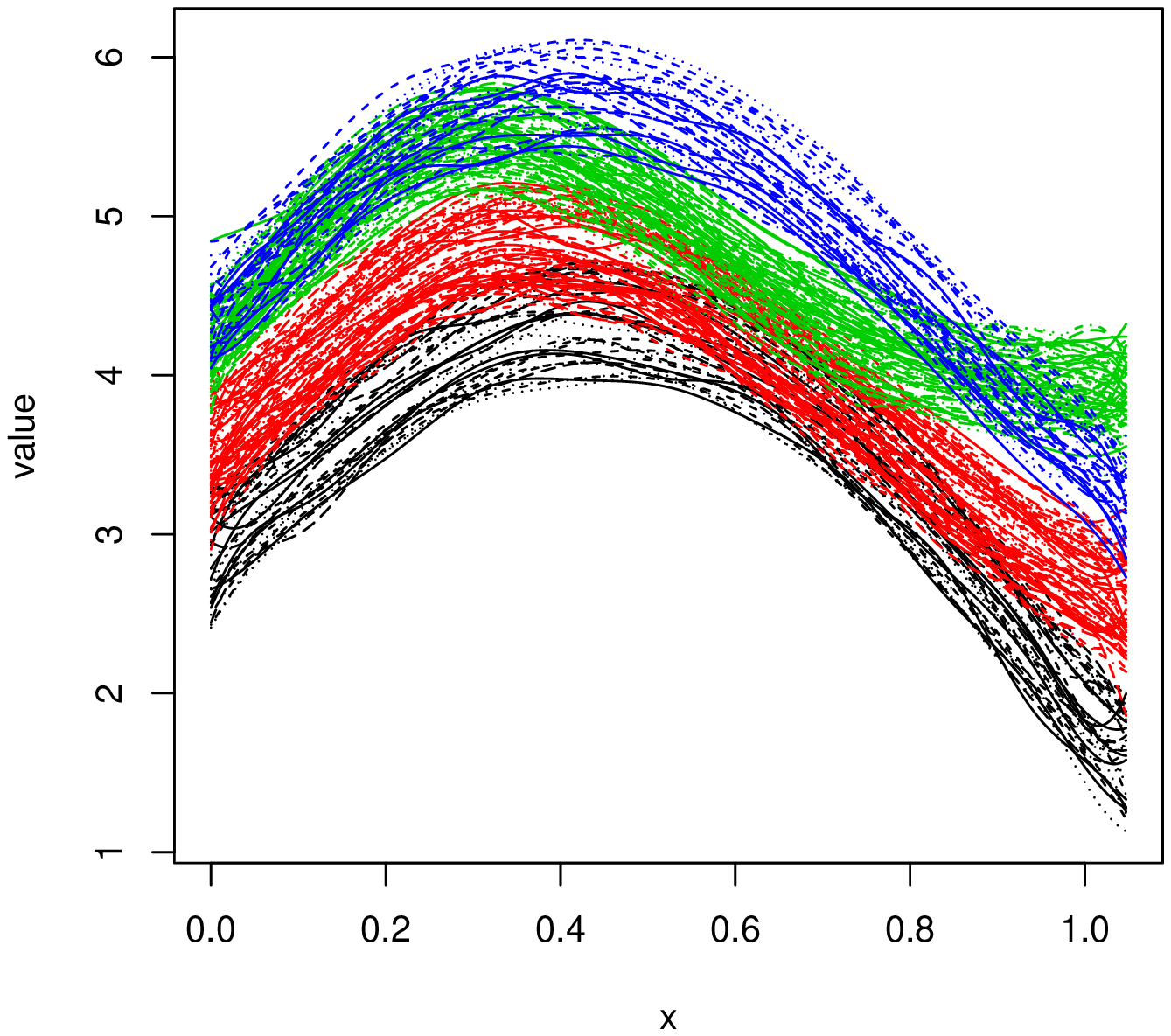}
        }
        \subfigure[]{%
            \label{fig:fourth}
            \includegraphics[width=5cm,height=3cm]{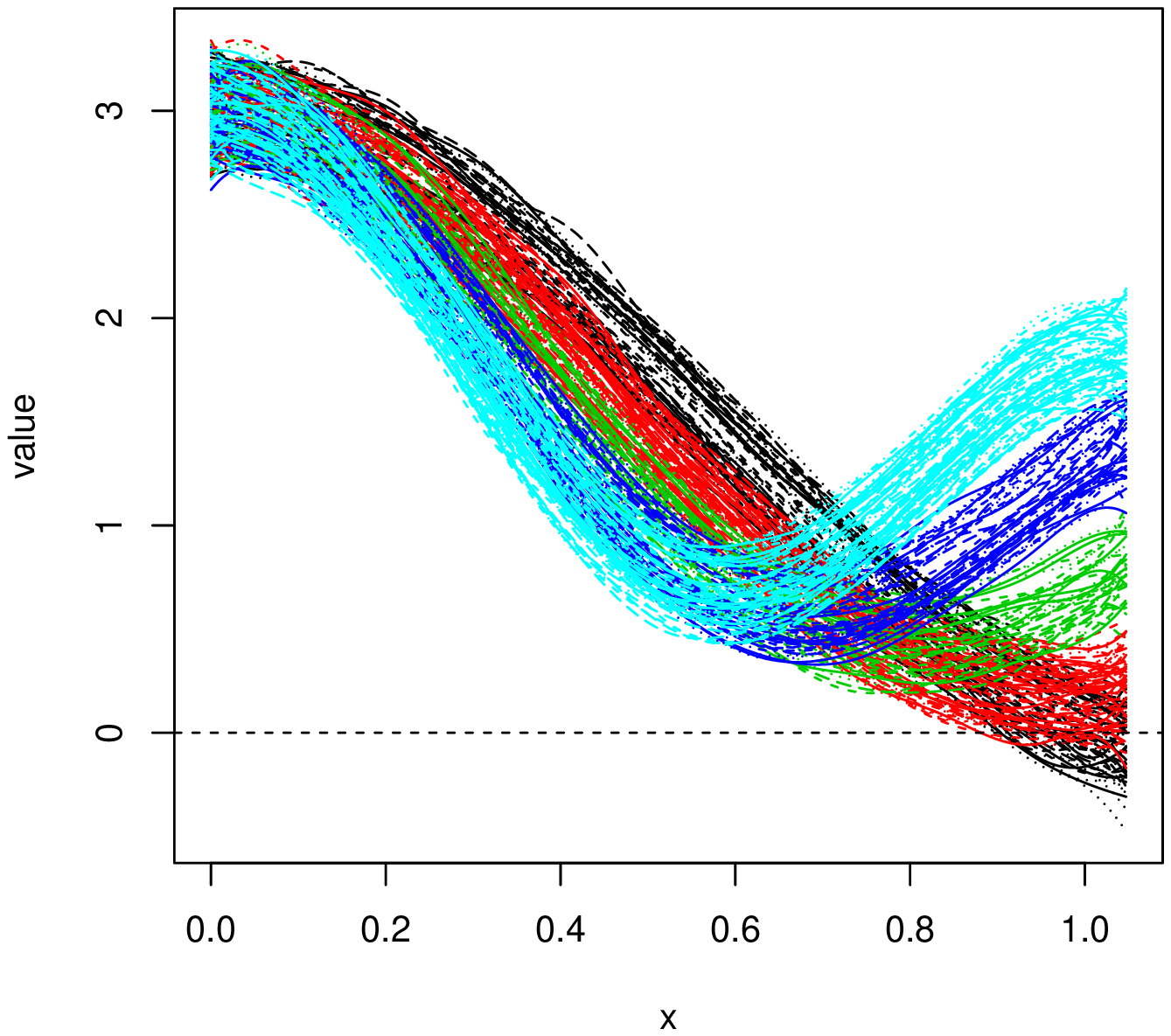}
        }%
    \end{center}
    \caption{%
        Functional data by clusters for (a) Scenario 1 ($K=2$), (b) Scenario 2 ($K=3$), (c) Scenario 3 ($K=4$) and (d) Scenario 4 ($K=5$).}
   \label{numclustsim}
\end{figure}

The smoothed versions from an example of the simulated curves for the 4 scenarios considered are shown in Figure \ref{numclustsim}, which present challenges not only to the selection of the number of clusters, but also to the identification of the clusters and the affiliation of each curve. The simulation results for each scenario in this section were based on 100 Monte Carlo runs while considering the possible number of clusters $k = 1, \ldots, 8$. For comparison purposes, the initial clustering for the $TW_\alpha$ and $L_2$ reported results were based on the K-means clustering algorithm in \cite{zambom2018functional}. Table \ref{tab.TW} shows the results of the simulations for Scenarios 1 to 4 when using the proposed $TW_\alpha$ measure with $\alpha$ chosen as described in Section \ref{select.alpha}, Table \ref{tab.L2} shows the results using $L_2$, and Table \ref{tab.Fun} shows the results of the methods specifically designed for functional data.

\begin{table}[!htb]
\footnotesize
\caption{Estimated number of clusters using methods in (\ref{CH}) - (\ref{DDSEJUMP}) with $TW_{\alpha}$ for scenarios 1 through 4.}
\label{tab.TW}
\centering
\begin{minipage}{\textwidth} 
\renewcommand{\thefootnote}{\thempfootnote} 
\begin{tabular*}{\textwidth}{@{\extracolsep{\fill}}lcccccccc} 
\toprule
& \multicolumn{8}{c}{Estimated number of clusters}\\
\cmidrule(lr){2-9}
Method & 1 & 2 & 3 & 4 & 5 & 6 & 7 & 8 \\
\midrule
\textbf{Scenario 1} ($K = 2$) & \\ 
CH 		 	&0 & 100 &   0 &   0  &  0  &  0  &  0  &  0 \\
KL 		 	&0 &  93 &   2  &  0 &   1  &  1  &  1  &  2\\
Hartigan 	&0 & 100 &   0  &  0 &   0  &  0  &  0  &  0\\
Silhouette 	&0 & 100 &   0  &  0 &   0  &  0  &  0  &  0\\
GAP 		&0 &  92 &   8  &  0 &   0  &  0  &  0  &  0\\
JUMP		&0 & 100 &   0  &  0 &   0  &  0  &  0  &  0\\
DDSE 		&0 &  65 &  8  & 23 &   3  &  1  &  0  &  0\\
Djump 		&0 &  81 &   6  &  4 &   8  &  1  &  0  &  0\\
\hline
\textbf{Scenario 2} ($K = 3$)& \\ 
CH 			&0 &  23  & 74 &   3  &  0  &  0  &  0 &   0\\
KL 			&0  &  0  & 92 &   5  &  3  &  0  &  0 &   0\\
Hartigan 	&0  & 27  & 56 &  16  &  1  &  0  &  0 &   0\\
Silhouette 	&0  &  5  & 95 &   0  &  0  &  0  &  0 &   0\\
GAP 		&29 &  43 &  28&    0 &   0 &   0 &   0&   0\\
JUMP 		&0  &  0  & 97 &   2  &  1  &  0  &  0 &   0\\
DDSE 		&8  & 14  & 75 &   1  &  2  &  0  &  0 &   0\\
Djump 		&0  &  0  & 85 &   7  &  5  &  3  &  0 &   0\\
\hline
\textbf{Scenario 3} ($K = 4$)& \\ 
CH 			&0  &  0  &100  &  0 &   0 &   0  &  0  &  0\\
KL 			&0  &  1  &  0  & 90 &   3 &   2  &  3  &  1\\
Hartigan 	&0  & 51  &  0  & 49 &   0 &   0  &  0  &  0\\
Silhouette 	&0  &100  &  0  &  0 &   0 &   0  &  0  &  0\\
GAP 		&19 &  63 &   0 &  18&    0&    0 &   0 &  0\\
JUMP 		&0  &  0 & 100  &  0 &   0 &   0  &  0  &  0\\
DDSE 		&0  &  9 &   3  & 85 &   2 &   1  &  0  &  0\\
Djump 		&0  &  0 &   1  & 84 &   5 &  10  &  0  &  0\\
\hline
\textbf{Scenario 4} ($K = 5$)& \\ 
CH 			&0  &  0 &   0  &  0 & 100  &  0  &  0  &  0\\
KL 			&0  &  0 &   0  &  0 & 100  &  0  &  0  &  0\\
Hartigan 	&0  &  0 &   0  &  0 & 100  &  0  &  0  &  0\\
Silhouette 	&0  &  0 &   0  &  0 & 100  &  0  &  0  &  0\\
GAP 		&0  &  3 &  19  & 34 &  39  &  5  &  0  &  0\\
JUMP 		&0  &  0 &   0  &  0 & 100  &  0  &  0  &  0\\
DDSE 		&0  &  0 &   0  &  2 &  96  &  1  &  1  &  0\\
Djump 		&1  &  4 &   0  &  1 &  90  &  4  &  0  &  0\\
\bottomrule
\end{tabular*}
\end{minipage} 
\end{table}

\newpage\clearpage

\begin{table}[!htb]
\footnotesize
\caption{Estimated number of clusters using methods in (\ref{CH}) - (\ref{DDSEJUMP}) with $L_{2}$  for scenarios 1 through 4.}
\label{tab.L2}
\centering
\begin{minipage}{\textwidth} 
\renewcommand{\thefootnote}{\thempfootnote} 
\begin{tabular*}{\textwidth}{@{\extracolsep{\fill}}lcccccccc} 
\toprule
& \multicolumn{8}{c}{Estimated number of clusters}\\
\cmidrule(lr){2-9}
Method & 1 & 2 & 3 & 4 & 5 & 6 & 7 & 8 \\
\midrule
\textbf{Scenario 1} ($K = 2$) & \\ 
CH 			&0 &  92  &  5 &   3  &  0 &   0  &  0  &  0\\
KL 			&0 &   9  &  6 &  26  & 19 &  16  & 13  & 11 \\
Hartigan 	&97&    0 &   2&    1 &   0&    0 &   0 &   0\\
Silhouette 	&0 & 100  &  0 &   0  &  0 &   0  &  0  &  0\\
GAP 		&24&    0&    7&    9 &  26&   33 &   1 &   0\\
JUMP 		&19&   29&    0&   20 &  20 &  10 &   1 &   1\\
DDSE 		&0 &   4 &   7 &  39  & 27  & 21  &  2  &  0\\
Djump 		&1 &  57 &  12 &  22  &  8  &  0 &   0  &  0\\
\midrule
\textbf{Scenario 2} ($K = 3$)& \\ 
CH 			&0  & 34  &  2  &  2   & 6  & 32  & 19  &  5\\
KL 			&0  & 11  &  2   & 6   & 8  & 24  & 22  & 27\\
Hartigan 	&63 &   1 &   1  &  5  &  5 &  10 &  10 &   2\\
Silhouette 	&0  & 98  &  0   & 0   & 0  &  2  &  0  &  0\\
GAP 		&76 &   0 &   2  &  5  &  6 &  10 &   1 &   0\\
JUMP 		&11 &   0 &   0  &  0  & 14  & 29 &  28 &  18\\
DDSE 		&2  & 47  &  7  & 12   & 8  & 21  &  3  &  0\\
Djump 		&25 &  57 &   7 &   7  &  3 &   1 &   0 &   0\\
\midrule
\textbf{Scenario 3} ($K = 4$)& \\ 
CH 			&0  &  0  &  0  & 99  &  0  &  1   & 0 &   0\\
KL 			&0  & 36  &  0  & 63  &  0  &  0   & 1 &   0\\
Hartigan 	&86 &   6 &   0 &   0 &   1 &   2  &  5&    0\\
Silhouette 	&0  &  1  &  0  & 99  &  0  &  0  &  0 &   0\\
GAP 		&0  & 47  &  0  &  5  & 25  & 21  &  2 &   0\\
JUMP 		&0  &  0  &  0 & 100  &  0  &  0  &  0 &   0\\
DDSE 		&0  &  0  &  0 &  77  & 11  &  6  &  6 &   0\\
Djump 		&90 &   0 &   0 &   9  &  1 &   0 &   0&    0\\
\midrule
\textbf{Scenario 4} ($K = 5$)& \\ 
CH 			&0 & 100 &   0  &  0   & 0  &  0   & 0  &  0\\
KL 			&0 &  76 &   0   & 0   & 0  &  5   &12  &  7\\
Hartigan 	&64 &   0&    0 &   0  &  0  & 16  & 19 &   1\\
Silhouette 	&0 & 100 &   0  &  0   & 0  &  0   & 0  &  0\\
GAP 		&0 &  29 &  46  & 23   & 2  &  0   & 0  &  0\\
JUMP 		&0 &  98 &   0  &  0   & 0  &  0  &  0  &  2\\
DDSE 		&0 &  17 &  46  & 27  &  7  &  3 &   0  &  0\\
Djump 		&7  & 28 &  51  & 10 &   2  &  2 &   0  &  0\\
\bottomrule
\end{tabular*}
\end{minipage} 
\end{table}


\begin{table}[!htb]
\footnotesize
\caption{Estimated number of clusters using methods designed for functional data.}
\label{tab.Fun}
\centering
\begin{minipage}{\textwidth} 
\renewcommand{\thefootnote}{\thempfootnote} 
\begin{tabular*}{\textwidth}{@{\extracolsep{\fill}}lcccccccc} 
\toprule
& \multicolumn{8}{c}{Estimated number of clusters}\\
\cmidrule(lr){2-9}
Method & 1 & 2 & 3 & 4 & 5 & 6 & 7 & 8 \\
\midrule
\textbf{Scenario 1} ($K = 2$)& \\ 
funclust    &26 &  44  &  24 &  5  &  1 &   0  &  0  &  0\\
funHDDC		&0 &   41  &  56 &  1  & 2 &  0  & 0  & 0 \\
curvclust 	&0&    0 &   0&    0 &   0&    8 &   37 &  5\\
funFEM  	&0&    0 &   0&    7 &  39&    46 &   8 &   0\\
MFclust		&0&    0 &   0&    0 &   0&    1 &   17 &   82\\
\midrule
\textbf{Scenario 2} ($K = 3$)& \\ 
funclust    &1 &  75  &  23 &  1  &  0 &   0  &  0  &  0\\
funHDDC		&0 &  63  &  28 &  1  &  2 &  3  &  2  &  1\\
curvclust 	&0 &  0  &  0 &  0  &  22&   7  &  8  &  63\\
funFEM  	&0 &  0  &  0 &  0  &  0 &   11  &  36  &  53\\
MFclust		&0 &  0  &  1&  0  &  1 &   7  &  21  &  71\\
\midrule
\textbf{Scenario 3} ($K = 4$)& \\ 
funclust    &0 &  62  &  37 &  1  &  0 &   0  &  0  &  0\\
funHDDC		&0 &  49  &  42 &  1  &  0 &   1  &  1  &  6\\
curvclust 	&0 &  0  &  0 &  0  &  0 &   10  &  54  &  36\\
funFEM  	&0 &  0  &  0 &  0  &  0 &   0  &  24  & 76\\
MFclust		&0 &  0  &  0 &  0  &  3 &   4  &  12  &  81\\
\midrule
\textbf{Scenario 4} ($K = 5$)& \\ 
funclust    &0 &  93  &  6 &  1  &  0 &   0  &  0  &  0\\
funHDDC		&0 &  2  &  14 &  0  &  7 &   0  &  5  &  72\\
curvclust 	&0 &  44  &  35 &  0  &  1 &   0  &  0  &  0\\
funFEM  	&0 &  0  &  0 &  0  &  0 &   0  &  0  &  100\\
MFclust		&0 &  0  &  0 &  0  &  1 &   5  &  31  &  63\\
\bottomrule
\end{tabular*}
\end{minipage} 
\end{table}

The results suggest that the proposed measure $TW_\alpha$, in general, improves the accuracy of all the modified procedures in selecting the number of clusters $K$ when compared to the respective results with the $L_2$ distance. When using $L_2$, most methods have high variability in the sense that the selected number of clusters frequently varies for different simulated datasets, even though the datasets come from the same underlying distributions. This high variability is probably due to the fact that the $L_2$ distance does not differ simultaneously the parallelism features and the mean difference of the curves, and thus the criteria used in each method fails to distinguish the correct optimal number of clusters. For example, Hartigan's and Gap's procedures using $L_2$ failed to select the correct number of clusters almost always in all simulation scenarios, while KL, JUMP, DDSE, and Djump only perform well in 1 out of the 4 scenarios. On the other hand, the results reported in Table \ref{tab.TW}  suggest that the proposed measure $TW_{\alpha}$ induces more stability to the selection of $K$ in the sense that throughout the simulated datasets, each method seems to frequently select the same optimal number of clusters, which is in the majority of cases correct. Only a few notable deviations from the correct number of clusters can be observed with the $TW_{\alpha}$ measure. The Gap criterium on Scenario 2 and 3 did not select the correct number of clusters in the majority of the simulations, and had high variability in Scenario 4. CH incorrectly selected $K$ in all simulations of Scenario 3, which also happened to JUMP and Silhouette. Finally, the results of the methods especially designed for functional data reported in Table \ref{tab.Fun} suggest that their performance is unsatisfactory for the challenging scenarios considered. Because their selection criterium is usually based on the BIC, it might have failed to identify the challenging characteristics of the curves by only computing the penalized likelihood.


\section{Real data analysis}\label{realdatanumclust5}

In this section we examine the effectiveness of using $TW_{\alpha}$ and $L_2$, as well as the functional methods, for the selection of the number of clusters on six real datasets: the Growth, the Moisture wheat, the Canadian weather, the Kneading, the Poblenou, and the Fat spectrum datasets. These datasets have been studied in the literature of clustering methods for functional data. In the first four datasets, the grouping is either known or artificially generated according to previous studies in the literature, while the last two datasets have no known groupings.

\begin{figure}[htbp!]
     \begin{center}
        \subfigure[]{%
            \label{fig:first}
            \includegraphics[width=5cm,height=3cm]{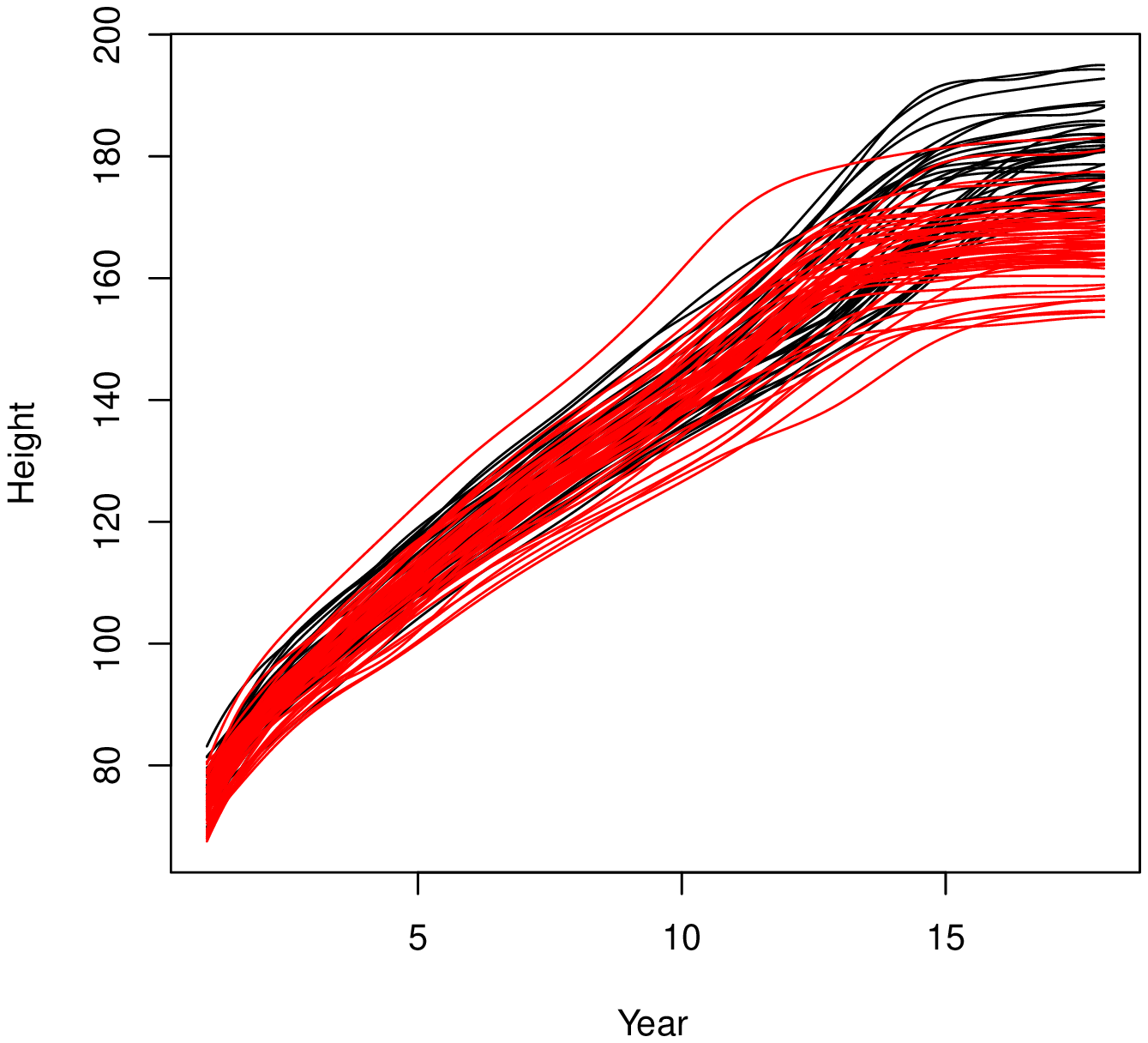}
        }
        \subfigure[]{%
            \label{fig:third}
            \includegraphics[width=5cm,height=3cm]{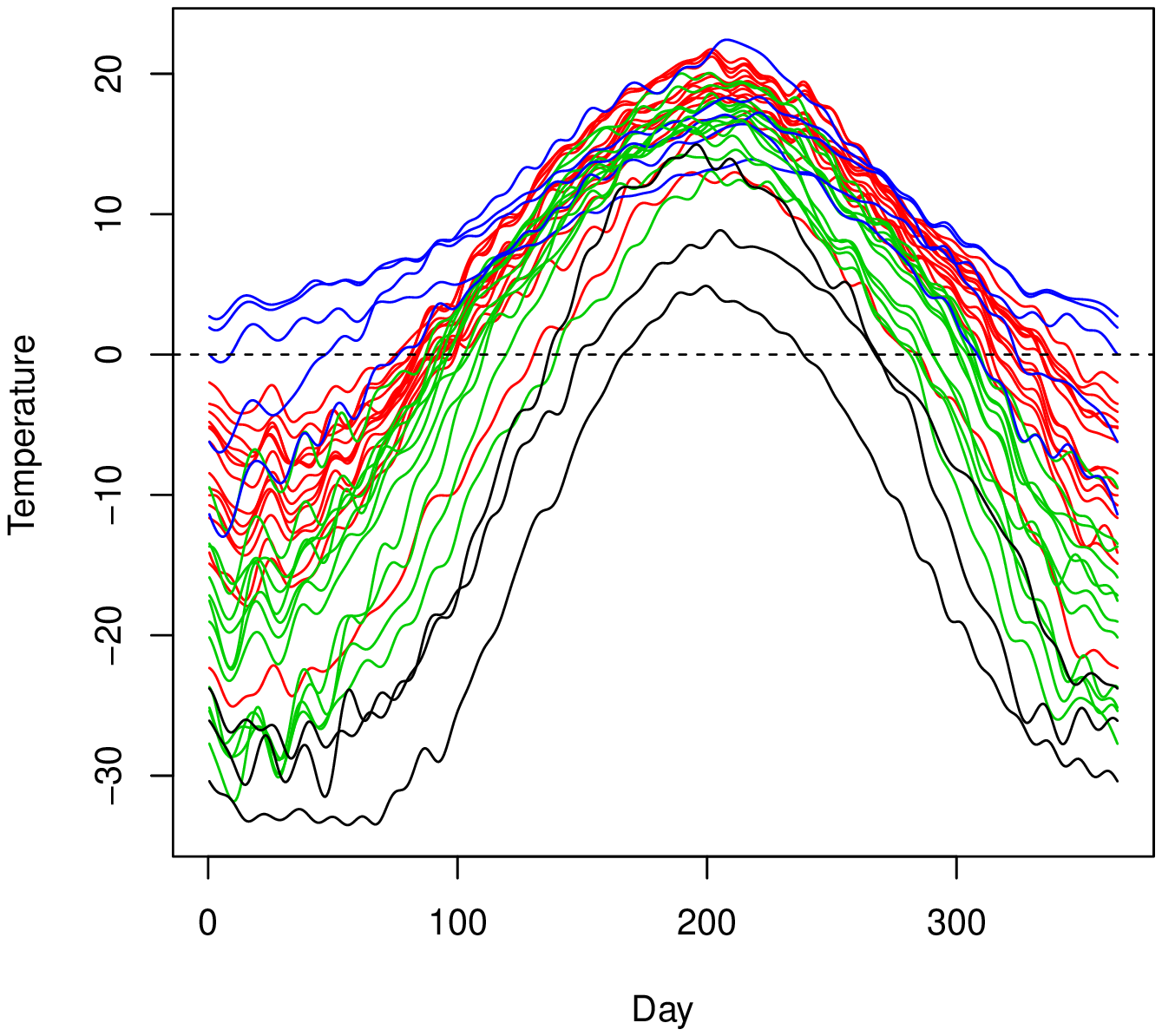}
        }
        \subfigure[]{%
            \label{fig:fourth}
            \includegraphics[width=5cm,height=3cm]{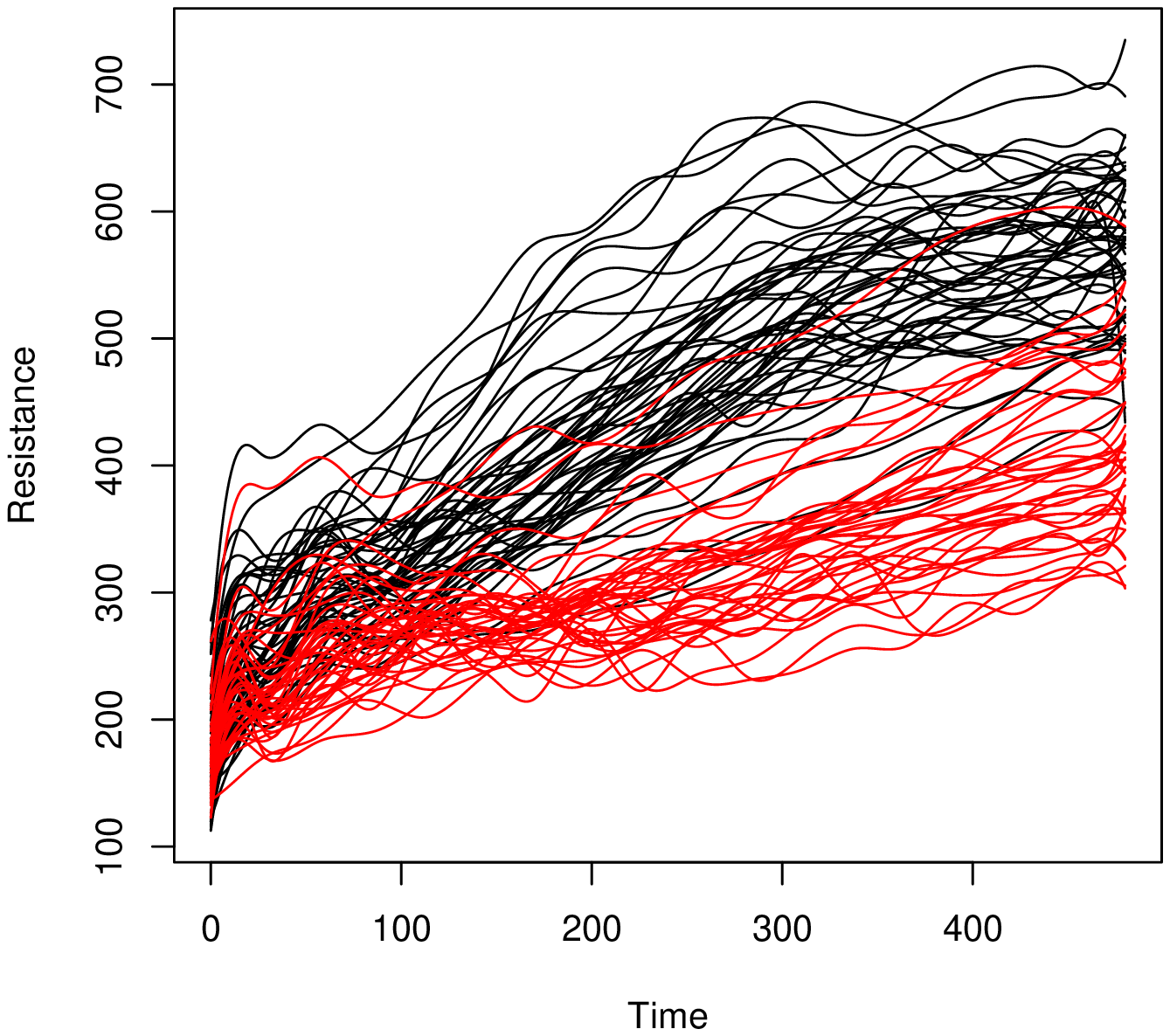}
        }%

    \end{center}
    \caption{%
        Functional datasets for first scenario: (a) Growth data, (b) Moisture wheat samples, (c) Canadian weather and (d) Kneading data.}
   \label{fig.DataKown}
\end{figure}

\begin{figure}[htbp!]
     \begin{center}
        \subfigure[]{%
            \label{fig:first}
            \includegraphics[width=5cm,height=3cm]{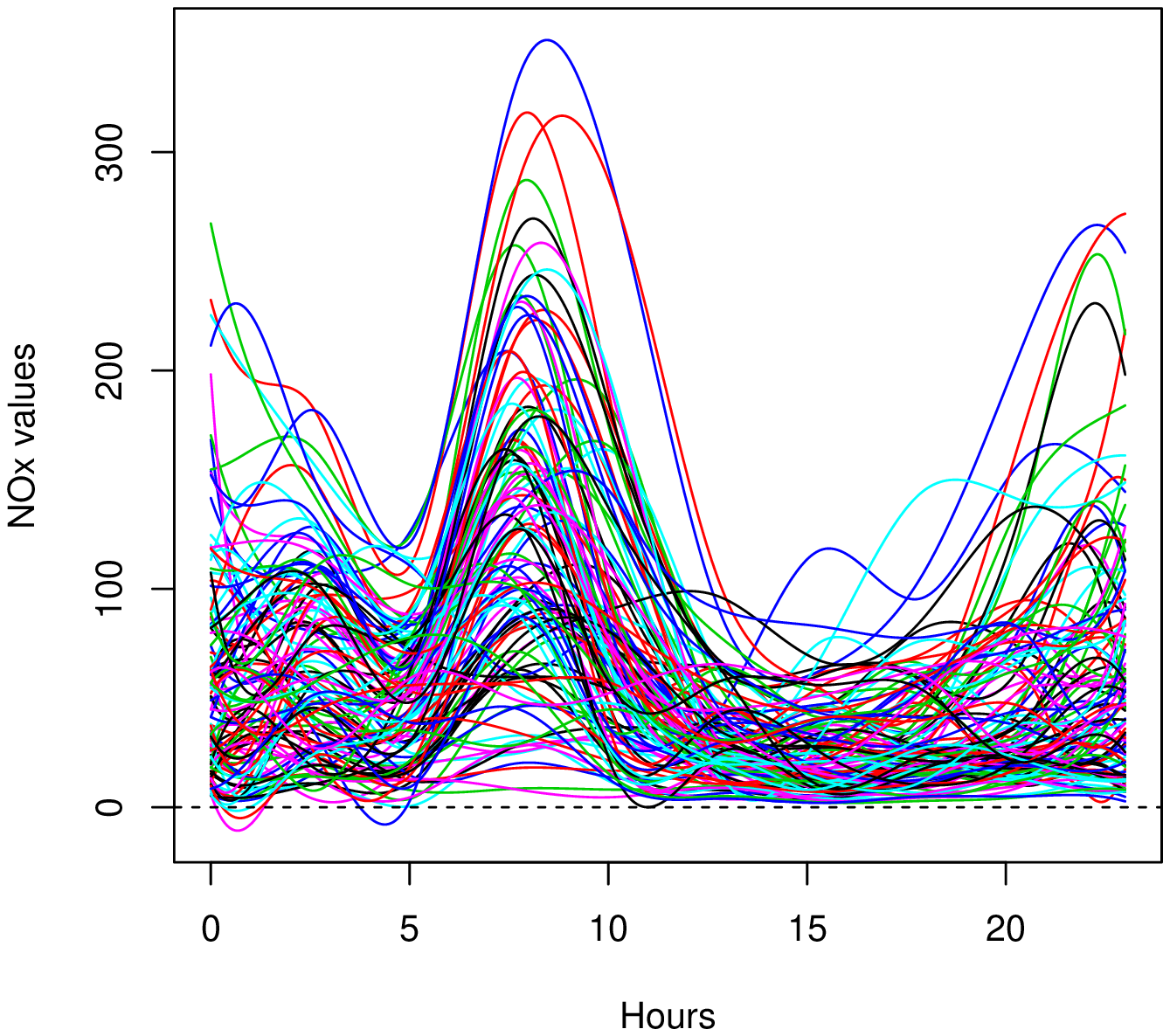}
        }
        \subfigure[]{%
           \label{fig:second}
           \includegraphics[width=5cm,height=3cm]{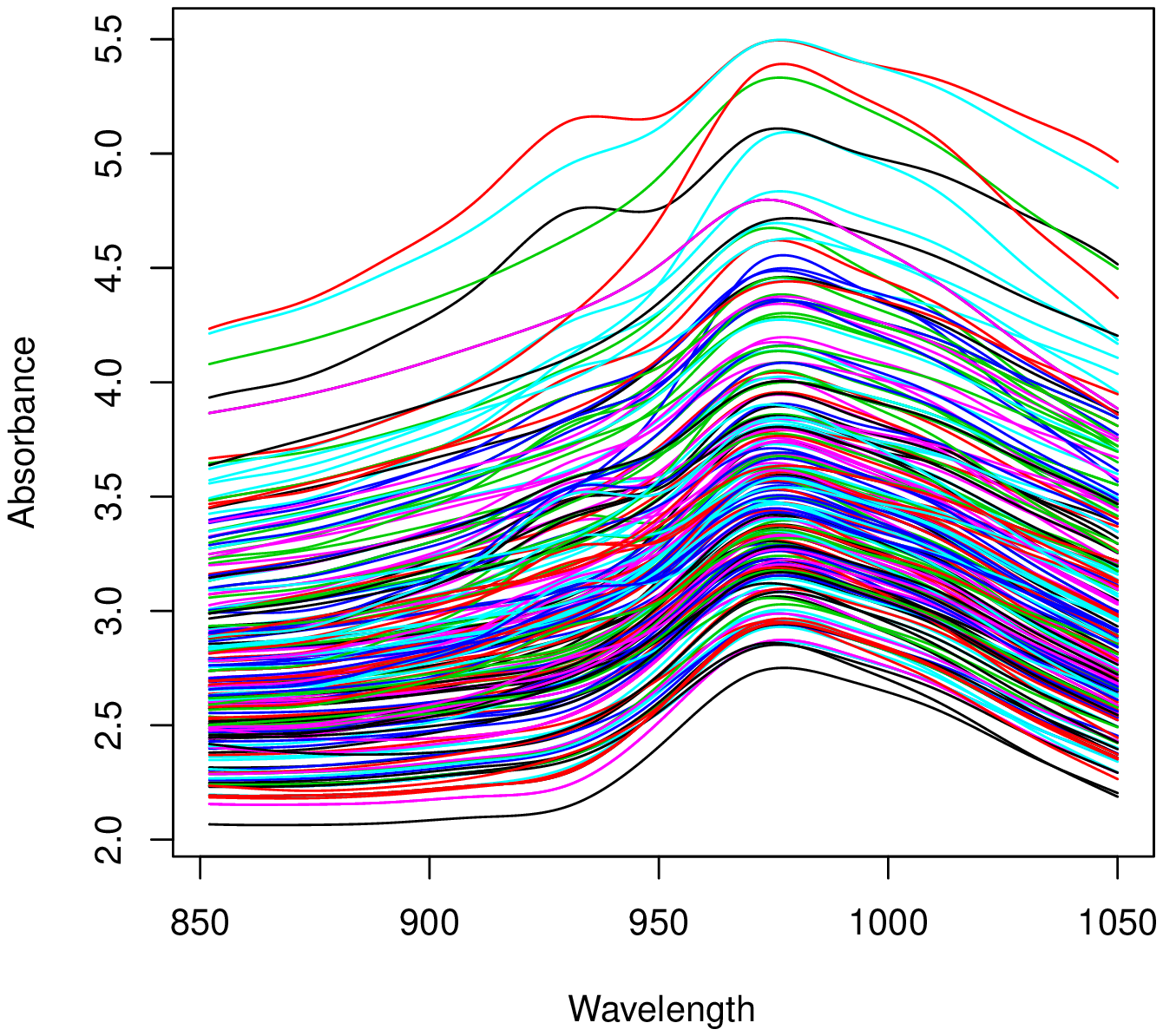}
        }

    \end{center}
    \caption{%
        Functional datasets for second scenario: (a) Poblenou data, (b) Fat spectrum data.}
   \label{fig.DataUnkown}
\end{figure}

The Growth dataset, a Berkeley longitudinal study, consists of the heights of 54 girls and 39 boys measured at 31 stages from 1 to 18 years of age. This dataset is  available in the R package \textit{fda}.
The Moisture wheat dataset consists of near-infrared reflectance spectra of 100 wheat samples, measured in 2 nm intervals from 1100 to 2500nm, along with the samples' moisture content. Since these samples had no natural grouping, we artificially split them into two groups as in Ferraty and Vieu's (section 8.4.2): assign the $n_1 = 41$ curves whose moisture content is smaller than 15 to group 1 and the remaining $n_2 = 59$ curves to group 2. This data was also analyzed in Kalivas (1997), and is available in R package \textit{fds}.
The Canadian weather dataset is also available in the R package \textit{fda} and consists of the daily temperature  
at 35 different weather stations in Canada averaged over 1960-1994; it is grouped in 4 regions (referring to climate zones Atlantic, Pacific, Continental, Arctic). 
The Kneading data consists of the recorded resistance of the dough during the kneading process over a time period of 480 seconds, with measurements at every 2 seconds. The data is publicly available at http://math.univ-lille1.fr/$\sim$preda/FDA/flours.txt. The first group is composed of 50 flours that yielded good quality cookies and 40 flours that yielded low quality cookies.
The last two datasets, whose groupings are not known, are: the Poblenou dataset, which measures the levels of NOx every hour by a control station in Poblenou, Barcelona, Spain (available in the R package \textit{fda.usc}); and the Fat spectrum dataset, which measures the absorbance at 100 wavelengths (from 852 to 1050 in step of 2nm) of fat content obtained by an analytic chemical processing (available in the R package \textit{fds}). The smoothed curves of the datasets considered are shown in Figures \ref{fig.DataKown} and \ref{fig.DataUnkown}.

Because the selection of the number of clusters depends on the clustering algorithm at each size $k = 1, \ldots, 8$, and the K-means clustering is initialized with random initial centers, the results shown in this section are based on 100 analyses of the same dataset, each of which has starting clusters randomly initialized for the K-means algorithm. 
The results of the analysis of the first four datasets with procedures using $TW_\alpha$ and $L_2$, and the ones designed for functional data are shown in Tables \ref{tab.TW.RealData}, \ref{tab.L2.RealData}, and \ref{tab.Fun.RealData}, respectively. The conclusions of this real data analysis are similar to those in the simulation studies in that the proposed modification of the methods using $TW_\alpha$ yields more stable selection of the optimal $K$, which is mostly accurate. The Gap method performed poorly with both $TW_\alpha$ and $L_2$, followed by DDSE and Djump, which presented high variability. All methods performed poorly in scenario 4, both with $TW_\alpha$ and $L_2$, probably due to the high proportion of the domain with overlapping curves. Interestingly, the methods designed for functional data performed poorly in all cases in selecting the number of clusters, as can be seen on Table \ref{tab.Fun.RealData}. Overall, the results suggest that proposed method  is more accurate in selecting the true number of clusters.

\begin{table}[!htb]
\footnotesize
\caption{Estimated number of clusters using methods in (\ref{CH}) - (\ref{DDSEJUMP}) with $TW_{\alpha}$ for the datasets: Growth, Moisture, Canadian Weather, and Kneading.}
\label{tab.TW.RealData}
\centering
\begin{minipage}{\textwidth} 
\renewcommand{\thefootnote}{\thempfootnote} 
\begin{tabular*}{\textwidth}{@{\extracolsep{\fill}}lcccccccccccccc} 
\toprule
& \multicolumn{8}{c}{Estimated number of clusters}\\
\cmidrule(lr){2-9}
Method & 1 & 2 & 3 & 4 & 5 & 6 & 7 & 8  \\
\midrule
\textbf{Growth} ($K = 2$)& \\ 
CH 		 	&0 & 100  &  0  &  0  &  0  &  0  &  0  &  0\\
KL 		 	&0 &  46  &  9  &  9  &  5  &  6  & 15 &  10\\
Hartigan 	&0 & 100  &  0  &  0  &  0  &  0 &   0  &  0\\
Silhouette 	&0 & 100  &  0  &  0  &  0  &  0  &  0  &  0\\
GAP 		&83&   17 &   0  &  0 &   0 &   0 &   0  &  0\\
JUMP		&0 & 100  &  0   & 0  &  0  &  0  &  0  &  0\\
DDSE 		&0 & 100  &  0   & 0  &  0  &  0  &  0  &  0\\
Djump 		&0 &  98  &  2   & 0 &   0  &  0  &  0  &  0\\
\hline
\textbf{Moisture} ($K = 2$)& \\ 
CH 			&0 & 100 &   0 &   0  &  0 &   0  &  0  &  0\\
KL 			&0 &  43  & 19 &   7 &  18 &   2  &  0 &  11\\
Hartigan 	&0 & 100 &   0 &   0  &  0 &   0  &  0 &   0\\
Silhouette 	&0 & 100  &  0 &   0  &  0  &  0  &  0  &  0\\
GAP 		&0 &  52 &  48  &  0  &  0  &  0 &   0 &   0\\
JUMP 		&0 & 100 &   0 &   0 &  0   & 0  &  0  &  0\\
DDSE 		&0 &  34 &  28 &  23 &   3  & 12  &  0 &   0\\
Djump 		&1 &  56 &  33 & 10  &  0  &  0 &   0  &  0\\
\hline
\textbf{Canadian Weather} ($K = 4$)& \\ 
CH 			&0 &  89  &  0 &  11  &  0 &   0 &   0  &  0\\
KL 			&0  &  0 &  12 &  24 &  29 &  17  & 13  &  5\\
Hartigan 	&0  & 94 &   6  &  0 &   0  &  0 &   0  &  0\\
Silhouette 	&0  & 92  &  8  &  0 &   0  &  0 &   0  &  0\\
GAP 		&85 &   0 &  15 &   0 &   0 &   0 &   0 &   0\\
JUMP 		&0  &  0  &  5  & 32 &  34  & 12  &  7  & 10\\
DDSE 		&0  &  1  & 17 &  42 &  29 &   8  &  3 &   0\\
Djump 		&3  & 28  & 21  & 25  & 22  &  1 &   0  &  0\\
\hline
\textbf{Kneading} ($K = 2$)& \\ 
CH 			&0 & 100  &  0  &  0  &  0 &   0  &  0 &   0\\
KL 			&0  & 81  &  5  &  2  &  8 &   0 &   3  &  1\\
Hartigan 	&0 & 100  &  0  &  0  &  0 &   0 &   0  &  0\\
Silhouette 	&0 & 100 &  0   & 0  &  0  &  0 &   0   & 0\\
GAP 		&0 &  89 &  11  &  0  &  0  &  0 &   0  &  0\\
JUMP 		&0 & 100  &  0  &  0  &  0  &  0  &  0  &  0\\
DDSE 		&0 &  80  &  1  & 19 &   0  &  0  &  0  &  0\\
Djump 		&0 &  69 &   5  & 23  &  3  &  0   & 0  &  0\\
\bottomrule
\end{tabular*}
\end{minipage} 
\end{table}

\begin{table}[!htb]
\footnotesize
\caption{Estimated number of clusters using methods in (\ref{CH}) - (\ref{DDSEJUMP}) with $L_{2}$ for the datasets: Growth, Moisture, Canadian Weather, and Kneading.}
\label{tab.L2.RealData}
\centering
\begin{minipage}{\textwidth} 
\renewcommand{\thefootnote}{\thempfootnote} 
\begin{tabular*}{\textwidth}{@{\extracolsep{\fill}}lcccccccc} 
\toprule
& \multicolumn{8}{c}{Estimated number of clusters}\\
\cmidrule(lr){2-9}
Method & 1 & 2 & 3 & 4 & 5 & 6 & 7 & 8 \\
\midrule
\textbf{Growth} ($K = 2$)& \\ 
CH 			&0 & 100  &  0 &  0 &   0  &  0  &  0  &  0\\
KL 			&0  & 80  &  0  &  8 &   0 &   6  &  2  &  4\\
Hartigan 	&70 &   30 &   0 &   0 &   0 &  0 &  0 &   0\\
Silhouette 	&0 & 100  &  0  &  0  &  0  &  0  &  0  &  0\\
GAP    		&100 &   0 &  0 &  0  & 0  &  0  &  0  &  0\\
JUMP 		&100 & 0  &  0  &  0  &  0  &  0  &  0  &  0\\
DDSE 		&0  &  82  &  0  &  11  & 5  &  2  &  0  &  0\\
Djump 		&0  &  63  &  0  &  35   &2  &  0  &  0  &  0\\
\hline
\textbf{Moisture} ($K = 2$)& \\ 
CH 			&0  &  100   & 0  &  0 & 0  &  0   & 0  &  0\\
KL 			&0  &  30   & 4  &  25  & 15  &  9   & 11  &  6\\
Hartigan 	&77  &  0  &  3 &   9 &  10 &  1  & 0 &   0\\
Silhouette 	&0   & 100  &  0  &  0 & 0  &  0  &  0  &  0\\
GAP 		&100   & 0  & 0  & 0 &  0  &  0  &  0  &  0 \\
JUMP 		&0   & 80  &  0  & 9 & 7  &  0  &  3  &  1\\
DDSE 		&0   & 3  &  25  &  41  & 23 &  4  &  4  &  0\\
Djump 		&2   &14 & 24 &  35 &  23  &  2  &  0  &  0\\
\hline
\textbf{Canadian Weather} ($K = 4$)& \\ 
CH 			&0 &  1 &   0  &  7 & 16 &   1 &   28  &  47\\
KL 			&0 &   4 &   7  &  7  & 45 &   2 &   24  &  11 \\
Hartigan 	&72&    4&   6  &  9 &  9&   0&    0 &   0\\
Silhouette 	&0 &   34 &   63   & 0 & 3 &   0 &   0  &  0\\
GAP 		&100 &   0 &  32  & 25  & 31 &  12 &   0  &  0\\
JUMP 		&0 &   0 &   0  &  6 & 26 &  2 &   38  &  28\\
DDSE 		&0 &   2 &   10  &  37 &  50 &   1 &   0  &  0\\
Djump 		&15 &  10 &   10  &  25 &  40 &   0 &   0  &  0\\
\hline
\textbf{Kneading} ($K = 2$)& \\ 
CH 			&0 &   100 &   0  &  0 & 100 &   0 &   0  &  0\\
KL 			&0 &   92 &   0  &  1  & 3 &   3 &   1  &  0 \\
Hartigan 	&66&    27&    5  &  1 &  1&   0&    0 &   0\\
Silhouette 	&0 &   100 &   0   & 0 & 100 &   0 &   0  &  0\\
GAP 		&0 &   98 &  2  & 0   & 0 &  0 &   0  &  0\\
JUMP 		&0 &   0 &   0  &  1 & 13 &   25 &  34 &  27\\
DDSE 		&0 &   87 &   0  &  1 &  3 &   7 &   1  &  1\\
Djump 		&8 &  89 &   0  &  2 &  5 &   1 &   2  &  1\\
\bottomrule
\end{tabular*}
\end{minipage} 
\end{table}

\begin{table}[!htb]
\footnotesize
\caption{Estimated number of clusters using methods designed for functional data for the datasets: Growth, Moisture, Canadian Weather, and Kneading..}
\label{tab.Fun.RealData}
\centering
\begin{minipage}{\textwidth} 
\renewcommand{\thefootnote}{\thempfootnote} 
\begin{tabular*}{\textwidth}{@{\extracolsep{\fill}}lcccccccc} 
\toprule
& \multicolumn{8}{c}{Estimated number of clusters}\\
\cmidrule(lr){2-9}
Method & 1 & 2 & 3 & 4 & 5 & 6 & 7 & 8 \\
\midrule
\textbf{Growth} ($K = 2$)& \\ 
funclust    &0 &  91  &  9 &   0  &  0  &  0  &  0 &   0\\
funHDDC		&0 & 100  &  0  &  0   & 0  &  0  &  0  &  0 \\
curvclust 	&0 &   0  &  0 & 100  &  0  &  0  &  0 &   0\\
funFEM  	&0 &   0  &  0 &   0  &  0  &  0  &  0 & 100\\
MFclust		&0  &  0 &   0  &  0  &  0  &  0  &  6  & 94\\
\midrule
\textbf{Moisture} ($K = 2$)& \\ 
funclust    &0  & 83  & 14&    3  &  0  &  0  &  0  &  0\\
funHDDC		&0  &  0  &  0 &   0 & 100  &  0 &   0  &  0\\
curvclust 	&0  &  0  &  0  &  0 &   1  &  5  & 32 &  62\\
funFEM  	&0  &  0  &  0  &  0 &   0  &  0  &  0 & 100\\
MFclust		&0  &  0 &   0  &  1 &   7  & 17 &  31  & 44\\
\midrule
\textbf{Canadian Weather} ($K = 4$)& \\ 
funclust    &0 & 95  &  5 &   0  &  0 &   0  &  0  &  0\\
funHDDC		&0 & 100 &   0  &  0  &  0 &   0  &  0  &  0\\
curvclust 	&0 &   0  &  0  &  0  &  0 &   0 &   3 &  97\\
funFEM  	&0  &  0  &  0  &100  &  0  &  0 &   0  &  0\\
MFclust		&0  &  0 &   0  &  0  &  5  &  5 &   0  & 90\\
\midrule
\textbf{Kneading} ($K = 2$)& \\ 
funclust    &0 &  65  & 25&    8  &  2 &   0  &  0 &   0\\
funHDDC		&0  &  0  &  0  &  0  &  0 &   0 & 100 &   0\\
curvclust 	&0  &  0  &  0 &   0  &  0 &   0  &  3 &  97\\
funFEM  	&0  &  0  &  0 &   0  &  0 & 100  &  0  &  0\\
MFclust		&0  &  0  &  0  &  0  &  0 &   3 &  20 &  77\\
\bottomrule
\end{tabular*}
\end{minipage} 
\end{table}

The results for the last two datasets, whose numbers of clusters are unknown, are shown in Tables \ref{tab.TW.RealUnknown}, \ref{tab.L2.RealUnknown}, and \ref{tab.Fun.RealUnknown}, for $TW_\alpha$, $L_2$, and functional methods, respectively. Although the real number of clusters is unknown, the proposed procedure yields a higher agreement across all the different methods, while the different procedures using $L_2$ and the functional methods estimate the optimal number of clusters with little agreement.

\begin{table}[!htb]
\footnotesize
\caption{Estimated number of clusters using methods in (\ref{CH}) - (\ref{DDSEJUMP}) with $TW_{\alpha}$ for the datasets: Poblenou and Fat Spectrum.}
\label{tab.TW.RealUnknown}
\centering
\begin{minipage}{\textwidth} 
\renewcommand{\thefootnote}{\thempfootnote} 
\begin{tabular*}{\textwidth}{@{\extracolsep{\fill}}lcccccccc} 
\toprule
& \multicolumn{8}{c}{Estimated number of clusters}\\
\cmidrule(lr){2-9}
Method & 1 & 2 & 3 & 4 & 5 & 6 & 7 & 8 \\
\midrule
\textbf{Poblenou} & & & & &\\ 
CH 		 	&0&  100  &  0 &   0  &  0  &  0  &  0 &   0\\
KL 		 	&0 &  13  &  5 &  17  &  9 & 14  & 30 & 12\\
Hartigan 	&0 & 100  &  0 &   0  &  0 &   0 &   0 &   0\\
Silhouette 	&0 & 100  &  0  &  0  &  0  &  0  &  0  &  0\\
GAP 		&0 & 100  &  0  &  0  &  0  &  0  &  0  &  0\\
JUMP		&0 & 100  &  0  &  0  &  0  &  0  &  0  &  0\\
DDSE 		&5 &  95  &  0  &  0 &   0  &  0  &  0  &  0\\
Djump 		&0 &  97 &   1  &  0 &   2  &  0  &  0  &  0\\
\hline
\textbf{Fat}& & & & &\\ 
CH 			&0 & 100  &  0 &   0  &  0  &  0  &  0 &   0\\
KL 			&0 &   0 &  68 &   5 &   0 &   2 &  10 &  15\\
Hartigan 	&0  &  0 & 100  &  0  &  0  &  0 &   0  &  0\\
Silhouette 	&0 & 100 &   0 &   0 &   0 &   0 &   0 &   0\\
GAP 		&0 &100 &   0  &  0  &  0  &  0  &  0  &  0\\
JUMP 		&0 & 100 &   0  &  0 &   0  &  0 &   0 &   0\\
DDSE 		&0 &   0 &  85  &  0 &   0  &  9 &   0  &  6\\
Djump 		&0 &  14 &  86  &  0 &  0  &  0  &  0  &  0\\
\bottomrule
\end{tabular*}
\end{minipage} 
\end{table}

\begin{table}[!htb]
\footnotesize
\caption{Estimated number of clusters using methods in (\ref{CH}) - (\ref{DDSEJUMP}) with $L_{2}$ for the datasets: Poblenou and Fat Spectrum}
\label{tab.L2.RealUnknown}
\centering
\begin{minipage}{\textwidth} 
\renewcommand{\thefootnote}{\thempfootnote} 
\begin{tabular*}{\textwidth}{@{\extracolsep{\fill}}lcccccccc} 
\toprule
& \multicolumn{8}{c}{Estimated number of clusters}\\
\cmidrule(lr){2-9}
Method & 1 & 2 & 3 & 4 & 5 & 6 & 7 & 8 \\
\midrule
\textbf{Poblenou} & & & & &\\ 
CH 			&0 & 100  &  0 &  0 &   0  &  0  &  0  &  0\\
KL 			&0  & 77  &  0  &  2 &   3 &  3  &  4  &  11\\
Hartigan 	&67 &   16 &   8 &   7 &   1 &  1 &  0 &   0\\
Silhouette 	&0 & 100  &  0  &  0  &  0  &  0  &  0  &  0\\
GAP    		&100 &   0 &  0 &  0  & 0  &  0  &  0  &  0\\
JUMP 		&4 & 0  &  0  &  0  &  1  &  15  &  46  &  34\\
DDSE 		&21  &  76  &  0  &  0  & 0  &  3  &  0  &  0\\
Djump 		&19  &  81  &  0  &  0   &0  &  0  &  0  &  0\\
\hline
\textbf{Fat}& & & & &\\ 
CH 			&0  &  0   & 0  &  0 & 0  &  11   & 42  &  47\\
KL 			&0  &  0   & 36  &  0  & 0  &  36   & 11  &  17\\
Hartigan 	&94  &  0  &  0 &   0 &  0 &  0  & 4 &  0\\
Silhouette 	&0   & 100  &  0  &  0 & 0  &  0  &  0  &  0\\
GAP 		&100   & 0  & 0  & 0 &  0  &  0  &  0  &  0 \\
JUMP 		&0   & 0  &  0  &  0 & 100  &  63  &  11  &  26\\
DDSE 		&49   & 0  &  0  &  0  & 0  &  51  &  0  &  0\\
Djump 		&100   &0  &  0  &  0 &  0  &  0  &  0  &  0\\
\bottomrule
\end{tabular*}
\end{minipage} 
\end{table}

\begin{table}[!htb]
\footnotesize
\caption{Estimated number of clusters using methods designed for functional data for the datasets: Poblenou and Fat Spectrum.}
\label{tab.Fun.RealUnknown}
\centering
\begin{minipage}{\textwidth} 
\renewcommand{\thefootnote}{\thempfootnote} 
\begin{tabular*}{\textwidth}{@{\extracolsep{\fill}}lcccccccc} 
\toprule
& \multicolumn{8}{c}{Estimated number of clusters}\\
\cmidrule(lr){2-9}
Method & 1 & 2 & 3 & 4 & 5 & 6 & 7 & 8 \\
\midrule
\textbf{Poblenou} & & & & &\\ 
funclust    &0  & 97 &   3  &  0 &   0 &   0  &  0 &   0\\
funHDDC		&0  &100 &   0  &  0&    0 &   0  &  0 &   0 \\
curvclust 	&0  &  0 &  97  &  3&    0 &   0  &  0 &   0\\
funFEM  	&0  &  0 &   0  &  0&  100 &   0  &  0 &   0\\
MFclust		&0  &  0 &   0  &  0&    0 &   0  &  2 &  98\\
\midrule
\textbf{Fat}& & & & &\\ 
funclust    &0 &  42 &  45  &  8  &  5 &   0  &  0 &   0\\
funHDDC		&0  &  0 &   0  &  0   & 0 &   0  &  0 & 100\\
curvclust 	&0  &  0 & 100  &  0  &  0  &  0  &  0 &   0\\
funFEM  	&0  &  0  &  0  &  0  &  0  &  0  &  0 & 100\\
MFclust		&0  &  0  &  0  &  0  &  0 &  55 &  26 &  19\\
\bottomrule
\end{tabular*}
\end{minipage} 
\end{table}

\newpage\clearpage
\section{Conclusion}\label{conclus5}

In this paper, we considered the problem of selection of the number of clusters for functional datasets. We proposed a new measure that combines two test statistics, one testing the difference in means of the curves and another testing their parallelism, which is inspired on the recent theory of high-dimensional one-way anova. The combination of these tests was used to modify the criteria in existing procedures, and select the optimal number of clusters. Results obtained in simulation studies illustrated the efficacy of the proposed measure compared to the use of $L_{2}$ and other methods specially designed for functional data analysis. Overall, the proposed modification improved the accuracy of procedures in selecting the correct number of clusters, as well as increased the agreement of the different procedures when compared to $L_2$. The illustration of the proposed method in real datasets also showed higher agreement across procedures when using the $TW_\alpha$ when compared to the use of $L_2$ or the functional methods. In general, the proposed approach combines a measure of parallelism and mean height difference between curves, which outperformed the competitors considered in both simulated and real data analysis considered in this paper.



\bibliographystyle{apalike} 
\bibliography{ClusRef.bib}   

\end{document}